\documentclass[twoside,twocolumn,11pt]{article}
\usepackage{setspace}
\usepackage{extsizes}
\usepackage[super,sort&compress,comma]{natbib} 
\usepackage[version=3]{mhchem}
\usepackage[left=1.5cm, right=1.5cm, top=1.785cm, bottom=2.0cm]{geometry}
\usepackage{balance}
\usepackage{times,mathptmx}
\usepackage{sectsty}
\usepackage{graphicx} 
\usepackage{lastpage}
\usepackage[format=plain,justification=justified,singlelinecheck=true,font={stretch=1.125,small,sf},labelfont=bf,labelsep=space]{caption}
\usepackage{float}
\usepackage{fancyhdr}
\usepackage{fnpos}
\usepackage[english]{babel}
\usepackage{array}
\usepackage{droidsans}
\usepackage{charter}
\usepackage[T1]{fontenc}
\usepackage[usenames,dvipsnames]{xcolor}
\usepackage{setspace}
\usepackage[compact]{titlesec}
\usepackage{changes}
\definechangesauthor[name={JK}, color=red]{JK}
\colorlet{Changes@Color}{red}

\definecolor{cream}{RGB}{222,217,201}

\begin{document}

\pagestyle{fancy}
\thispagestyle{plain}
\fancypagestyle{plain}{
\fancyhead[C]{\includegraphics[width=18.5cm]{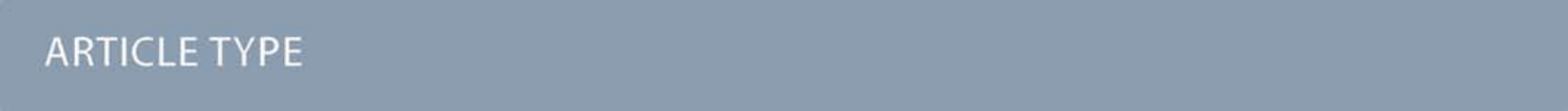}}
\fancyhead[L]{\hspace{0cm}\vspace{1.5cm}\includegraphics[height=30pt]{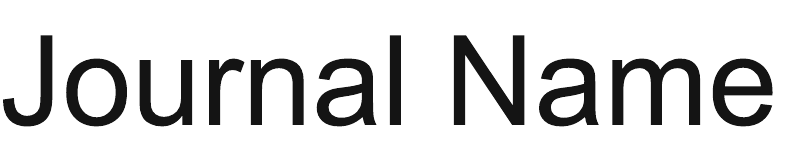}}
\fancyhead[R]{\hspace{0cm}\vspace{1.7cm}\includegraphics[height=55pt]{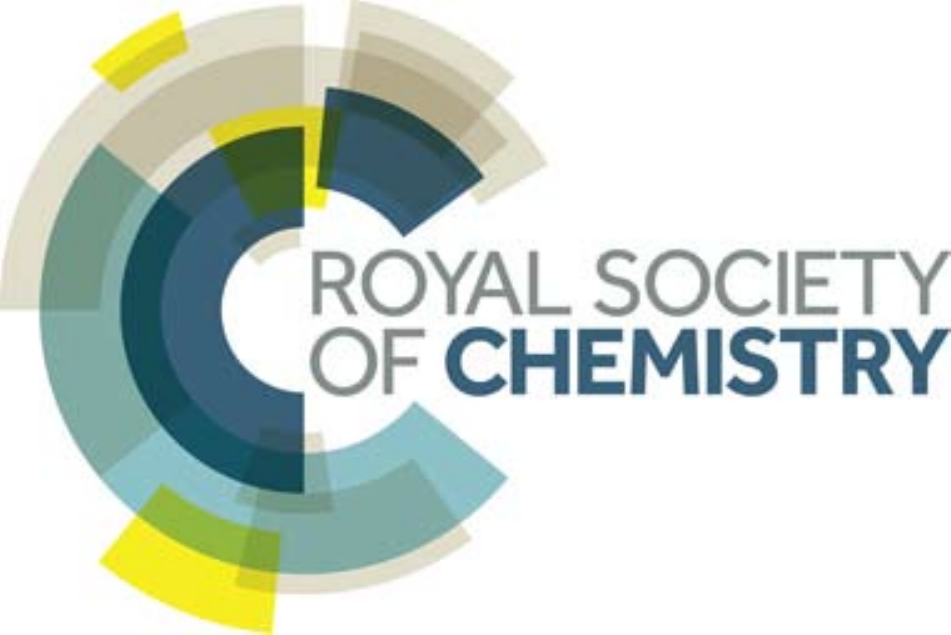}}
\renewcommand{\headrulewidth}{0pt}
}

\makeFNbottom
\makeatletter
\renewcommand\LARGE{\@setfontsize\LARGE{15pt}{17}}
\renewcommand\Large{\@setfontsize\Large{12pt}{14}}
\renewcommand\large{\@setfontsize\large{10pt}{12}}
\renewcommand\footnotesize{\@setfontsize\footnotesize{7pt}{10}}
\makeatother

\renewcommand{\thefootnote}{\fnsymbol{footnote}}
\renewcommand\footnoterule{\vspace*{1pt}%
\color{cream}\hrule width 3.5in height 0.4pt \color{black}\vspace*{5pt}} 
\setcounter{secnumdepth}{5}

\makeatletter 
\renewcommand\@biblabel[1]{#1}            
\renewcommand\@makefntext[1]%
{\noindent\makebox[0pt][r]{\@thefnmark\,}#1}
\makeatother 
\renewcommand{\figurename}{\small{Fig.}~}
\sectionfont{\sffamily\Large}
\subsectionfont{\normalsize}
\subsubsectionfont{\bf}
\setstretch{1.125} 
\setlength{\skip\footins}{0.8cm}
\setlength{\footnotesep}{0.25cm}
\setlength{\jot}{10pt}
\titlespacing*{\section}{0pt}{4pt}{4pt}
\titlespacing*{\subsection}{0pt}{15pt}{1pt}

\fancyfoot{}
\fancyfoot[LO,RE]{\vspace{-7.1pt}\includegraphics[height=9pt]{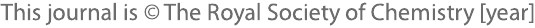}}
\fancyfoot[CO]{\vspace{-7.1pt}\hspace{13.2cm}\includegraphics{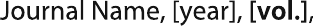}}
\fancyfoot[CE]{\vspace{-7.2pt}\hspace{-14.2cm}\includegraphics{RF}}
\fancyfoot[RO]{\footnotesize{\sffamily{1--\pageref{LastPage} ~\textbar  \hspace{2pt}\thepage}}}
\fancyfoot[LE]{\footnotesize{\sffamily{\thepage~\textbar\hspace{3.45cm} 1--\pageref{LastPage}}}}
\fancyhead{}
\renewcommand{\headrulewidth}{0pt} 
\renewcommand{\footrulewidth}{0pt}
\setlength{\arrayrulewidth}{1pt}
\setlength{\columnsep}{6.5mm}
\setlength\bibsep{1pt}

\makeatletter 
\newlength{\figrulesep} 
\setlength{\figrulesep}{0.5\textfloatsep} 

\newcommand{\topfigrule}{\vspace*{-1pt}%
\noindent{\color{cream}\rule[-\figrulesep]{\columnwidth}{1.5pt}} }

\newcommand{\botfigrule}{\vspace*{-2pt}%
\noindent{\color{cream}\rule[\figrulesep]{\columnwidth}{1.5pt}} }

\newcommand{\dblfigrule}{\vspace*{-1pt}%
\noindent{\color{cream}\rule[-\figrulesep]{\textwidth}{1.5pt}} }

\makeatother

\twocolumn[
  \begin{@twocolumnfalse}
\vspace{3cm}
\sffamily
\begin{tabular}{m{4.5cm} p{13.5cm} }

\includegraphics{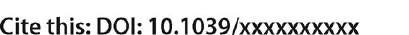} & \noindent\LARGE{\textbf{Non-affine lattice dynamics of defective fcc crystals$^\dag$}} \\
\vspace{0.3cm} & \vspace{0.3cm} \\

 & \noindent\large{Johannes Krausser\textit{$^{a}$} Rico Milkus\textit{$^{a}$} and Alessio Zaccone\textit{$^{a,b}$}} \\

\includegraphics{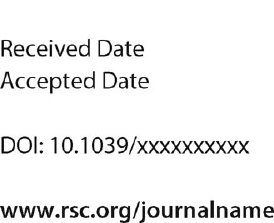} & \noindent\normalsize{
The mechanical, thermal and vibrational properties of defective crystals are important in many different contexts, from metallurgy and solid-state physics to, more recently, soft matter and colloidal physics. Here we study two different models of disordered fcc crystal lattices, with randomly-removed bonds and with vacancies, respectively, within the framework of non-affine lattice dynamics. 
We find that both systems feature the same scaling of the shear modulus with the newly defined inversion-symmetry breaking (ISB) parameter, which shows that local inversion-symmetry breaking around defects is the universal root source of the non-affine softening of the shear modulus. This finding allows us to derive analytical relations for the non-affine (zero-frequency) shear modulus as a function of vacancy concentration in excellent agreement with numerical simulations.
Nevertheless, due to the different microstructural disorder, the spatial fluctuations of the local ISB parameter are different in the vacancy and bond-depleted case. The vacancy fcc exhibits comparatively a more heterogenous microstructural disorder (due to the broader distribution of coordination number $Z$), which is reflected in a different scaling relation between boson peak frequency in the DOS and the average $\bar{Z}$.
These differences are less important at low vacancy concentrations, where the numerical DOS of the vacancy fcc can be well described theoretically by coherent-potential approximation, presented here for the bond-depleted fcc lattice in 3d. 
} \\

\end{tabular}

 \end{@twocolumnfalse} \vspace{0.6cm}

  ]

\renewcommand*\rmdefault{bch}\normalfont\upshape
\rmfamily
\section*{}
\vspace{-1cm}


\footnotetext{\textit{$^{a}$~Statistical Physics Group, Department of Chemical Engineering and Biotechnology, University of Cambridge, CB2 3RA Cambridge, U.K.}}
\footnotetext{\textit{$^{b}$~Cavendish Laboratory, University of Cambridge, CB3 0HE }}





%
%
%
%
%
%
%
%

%
%
%
%
%
\section{Introduction}
Understanding the mechanical properties of disordered materials at the level of their constituent building blocks has widespread applications, from metallurgy to relatively new fields such as photonic materials~\cite{Braun2006,Zhao2016}.
To achieve this goal, it is important to identify suitable model systems where the atomic-scale or particle-scale physics can be described by means of theoretical frameworks to provide sufficiently general principles and insights.
From this point of view, fcc crystals with point defects represent an ideal model system: they are amenable to theoretical approaches and at the same time are found in important technological applications. For example, photonic crystals made of colloidal particles are intensely studied for the opportunity they offer to control and manipulate light flow through a material, where the photonic band gap can be tuned by the particle size and lattice spacing~\cite{Imhof2003,Lidorikis2004}. Stable colloidal crystal phases are most of the time fcc lattices with point defects, mostly vacancies, and are promising materials also for optical computing. 

In a completely different setting, plutonium in its $\delta$-phase presents fcc structure in a temperature range that is important for energy applications, and one of the long-standing unsolved problems in plutonium metallurgy is to understand the mechanism of ageing induced by self-irradiation: this is most likely to involve formation of vacancies which coalesce into voids filled with He atoms from $\alpha$-decay processes. Clearly, it is therefore important to have an atomic-level mechanistic understanding of the mechanical stability of plutonium in its fcc phase and of e.g. the dependence of the shear modulus on the vacancy concentration~\cite{Hecker2000, Schwartz2007}.

In spite of a large body of theoretical and experimental literature across different fields, the possibility of explaining the puzzling properties of these disordered crystalline systems within a single framework and, crucially, with a single parameter that embodies the relevant particle-scale symmetry or lack thereof, has remained elusive. Pioneering theoretical work by I.M. Lifshitz~\cite{Kosevich2005} on the lattice dynamics of crystals with point defects has introduced elegant Green's functions methods which are accurate in the single-defect limit but difficult to extend to higher defect concentrations.

Here we aim to provide such a unifying framework to connect the mechanical and vibrational properties with the microstructure and its symmetries using the non-affine response formalism together with the key concept of local inversion symmetry. 
We will focus on two types  of disordered fcc lattices: i) fcc lattices with randomly-removed bonds -- what we call depleted fcc; ii) fcc lattices with vacancies.
The latter are of course directly relevant to applications, as discussed above, but at the same time less amenable to theoretical analysis. Hence we aim to determine under which conditions and in what range of parameters the two systems can be described by the same theory in a unified way. 
In our analysis, both numerical and theoretical, we assume that atoms/particles on the lattice are interacting harmonically with nearest-neigbhours only.
Furthermore, we neglect thermal fluctuations and focus on the limit of low-temperature or athermal solids which allows us to more clearly disentangle the complex relationship between local lattice symmetry and emerging elastic and vibrational properties. Furthermore, the low-temperature limit, where the control parameter is the atomic/particle packing fraction (or its lattice analogue: the coordination number), instead of temperature, is directly relevant to the case of colloidal crystals. 

In the following, we start by recalling the core concepts of the non-affine lattice dynamics framework. We then apply it to compare the two  types of disordered fcc lattices with regard to their shear elasticity as a function of defect concentration. Then we consider the vibrational density of states for the two systems, both by numerical diagonalizing the underlying Hessian of the configurations and by means of a coherent potential approximation.

\section{Non-affine lattice dynamics}

In order to study the differences and similarities of the vibrational and elastic properties of the bond-depleted fcc and of the fcc with vacancies, we shortly introduce the framework of non-affine lattice dynamics~\cite{Lemaitre2006}.
 
The starting point of non-affine lattice dynamics is the realization that in a disordered solid
the standard affine approximation of the Born-Huang theory~\cite{Born1988} of lattice dynamics breaks down. In other words, applying a shear strain $\gamma$ to a disordered solid leads, in addition to the affine displacement of the particles, which is directly proportional to the applied strain, to a non-affine contribution to the displacement fields.
For example, if the original position of a test particle in the undeformed lattice is $\underline{R}_{0}$, the affine position after a shear deformation is defined as $\underline{r}_{i,A}=\underline{\underline{\eta}} \cdot\underline{R}_{i,0}$, where $\underline{\underline{\eta}}$ is the strain tensor. 

In a perfectly centrosymmetric lattice, the particle en route towards this affine position receives forces from its nearest-neighbors which cancel each other out by symmetry, leaving the particle at equilibrium in the affine position. In a disordered lattice, however, due to the \textit{local breaking of inversion symmetry} on the given particle, these forces do not cancel, and their vector sum yields a net force that brings the particle to a final (\textit{non-affine}) position which differs from $\underline{r}_{i,A}$.  

These forces, which bring the particle away from the affine position into the final non-affine position, can be written out within the harmonic approximation of the interaction potential $V$, as 
  $\underline{f}_i = \underline{\Xi}_i \gamma$. In the absence of pre-stress (i.e. all bonds are relaxed in the harmonic energy minimum), it can be shown~\cite{Lemaitre2006} that $\underline{\Xi}_i = - \kappa r_0 \sum_j \underline{\hat{n}}_{ij} n^x_{ij} n^y_{ij}$, where $\underline{\hat{n}}_{ij}$ is the unit vector that connects two particles $i$ and $j$ on the lattice.

Furthermore, the internal work which is required to displace the particles from their (virtual) affine positions to their non-affine (final equilibrium) positions contributes negatively to the free energy of deformation, i.e.
\begin{align}
		F(\gamma) = F_{\text{A}}(\gamma)-F_{\text{NA}}(\gamma).
\end{align}

%

The shear modulus $G$ of a disordered solid can thus be derived from the free energy of deformation using $G = \partial^2 F / \partial \gamma^2$ and is given by
\begin{align}
	G = G_{\text{A}} - G_{\text{NA}}
	=
	\dfrac{1}{V} \left( \dfrac{\partial^2 U}{ \partial \gamma^2} \bigg \vert_{\gamma \to 0}
	- \underline{\Xi}\; \cdot \underline{\underline{H}}^{-1}\cdot \underline{\Xi},
\right)
\label{eq_G}
\end{align}
where the components of the Hessian $\underline{\underline{H}}$ are given by ${ H}_{ij} = \frac{\partial^2 U}{ \partial \underline{r}_i\partial  \underline{r}_i} \big \vert_{\gamma \to 0}$. 
As stated above, The affine force vector is defined as
 $\underline{\Xi} = - \dfrac{\partial^2 U}{ \partial \underline{r}_i\partial \gamma} \bigg \vert_{\gamma \to 0}
	 =
	 \kappa r_0 \sum_j n_{ij}^x n_{ij}^y \hat{n}_{ij}$.
The affine part of the above shear modulus is the standard Born-Huang formula, i.e.
\begin{align}
G_{\text{A}} 
= \dfrac{1}{V}\dfrac{\partial^2 U}{ \partial \gamma^2} \bigg \vert_{\gamma \to 0}
=\dfrac{\kappa r_0^2}{2 V} \sum_{ij} \left( n_{ij}^x n^y_{ij} \right)^2.
\label{eq_G_A}
\end{align}
Since the non-affine contribution to the shear modulus is proportional to the vector $\underline{\Xi}_i $, it vanishes for a perfect centrosymmetric crystal: the sum of triplets $\sum_j \underline{\hat{n}}_{ij} n^x_{ij} n^y_{ij}$ is identically zero if the nearest-neighbours are arranged symmetrically around the atom $i$, as one can easily verify. In other words, this is a consequence of the fact that the affine force field $\underline{\Xi}_{i}$ is non-zero if and only if the local inversion symmetry is broken.

\subsection{Shear modulus of the ideal and bond-depleted fcc crystals}
It will be instructive to use the above formalism and derive the shear modulus for a perfect three-dimensional fcc crystal. 
Realizing that under the application of a pure $x$-$y$ shear strain $\gamma$ only the four bonds lying in the $x$-$y$-plane contribute to the shear modulus $G$, we simply have to use Eq.~\eqref{eq_G_A} and evaluate the sum appearing there. The four bond vectors in the $x$-$y$-plane which contribute to this sum are given by  $\hat{\underline{n}}_1 = (1,1,0)^T/\sqrt{2}$, $\hat{\underline{n}}_2 = (-1,1,0)^T/\sqrt{2}$, $
\hat{\underline{n}}_3 = (1,-1,0)^T/\sqrt{2}$ and  
$\hat{\underline{n}}_4 = (-1,-1,0)^T/\sqrt{2}$.
Consequently, by virtue of Eq.~(\ref{eq_G_A}) the shear modulus of the perfect fcc crystal is given by
\begin{align}
G_{\text{fcc}} = \dfrac{1}{2} \rho r_0^2 \kappa = \dfrac{\kappa}{a}
\end{align}
using $\rho r_0^2 = 2/a$, where $a$ is the lattice constant of the fcc crystal, $\rho = N/V$ and $r_0$ the equilibrium bond length.
We can extend this result to the case of a depleted fcc crystal, where bonds are randomly cut 
by evaluating the number of bonds that contribute on average to the $x$-$y$-plane as
\begin{align}
G_{\text{A}}^{\text{depl}}(\bar{Z})
=
\sum_{i=0}^4 i \dfrac{\kappa}{4a}\dfrac{{{4}\choose{i}}	{{8}\choose{\bar{Z}-i}}	}{ {{12}\choose{\bar{Z}}}}
=
\dfrac{\kappa}{a}\dfrac{\bar{Z}}{12},
\end{align}
where $\bar{Z} \equiv \langle  Z \rangle= \frac{1}{N}\sum_{i=1}^N Z_i P(Z_i)$ is the disorder-averaged connectivity.
$P(Z_i)$ denotes the distribution of the local coordination numbers $Z_i$.
In the above equation, the expression $\frac{{{4}\choose{i}}	{{8}\choose{\bar{Z}-i}}	}{ {{12}\choose{\bar{Z}}}}$ represents the probability of having $i$ bonds in the $x$-$y$-plane which depends on the total number of nearest-neighbor bonds $\bar{Z}$.

The shear modulus of a depleted fcc crystal vanishes at $\bar{Z}=6$, the isostatic point of marginal stability, as described by Maxwell counting~\cite{Thorpe2002}.
Using this fact, we have that $G(\bar{Z}=6)=G_{\text{A}}(\bar{Z}=6) - G_{\text{NA}}(\bar{Z}=6)=0$. The non-affine contribution exactly cancels the affine modulus at $\bar{Z}=6$, which is a common feature of random central-force lattices~\cite{Zaccone2011}. In addition to that, $G_{\text{NA}}$ should vanish for $\bar{Z}=12$, i.e. for the case of an ideal fcc crystal the non-affine softening is absent. Assuming the linear behavior of the non-affine contribution\cite{Milkus2016}, which is justified a posteriori through numerical simulations, $G_{\text{NA}}$ can be written as the interpolation
between the above two cases at $\bar{Z}=6$ and $\bar{Z}=12$, i.e.
\begin{align}
G_{\text{NA}}^{\text{depl}}(\bar{Z}) = \dfrac{\kappa}{a}\dfrac{12-\bar{Z}}{12}.
\end{align}
Subtracting the non-affine contribution from the affine shear modulus we arrive at a simple expression for the shear modulus of the bond-depleted fcc crystal~\cite{Milkus2016}
\begin{align}\label{eq_shear_depl}
	G^{\text{depl}}(\bar{Z}) &= G_{\text{A}}^{\text{depl}}(\bar{Z})-G_{\text{NA}}^{\text{depl}}(\bar{Z})	
\\
	&= \dfrac{\kappa}{a}\dfrac{\bar{Z}-6}{6}.
\end{align}

\section{From the bond-depleted to  the vacancy fcc}
When considering bond-depleted systems it is natural to use the bond-occupation probability $p$ as a control parameter. It is given by
$p=\bar{Z}/Z_0$, where $Z_0=12$ is the number of nearest-neighbour bonds in the perfect fcc crystal.

We now establish a connection between the physical properties of the bond-depleted case and a fcc crystal with vacancies where a certain fraction of particles is removed thereby introducing defects.

By randomly removing $N'$ particles from the $N$ lattice sites, we are left with $N-N'$ particles. In this sense, we want to describe the elastic and vibrational properties of a fcc crystal with vacancies as a function of the vacancy concentration $c=N'/N$ with a new particle density $\rho'= (N-N')/V$.

As we will see, it is possible to obtain the average connectivity $\bar{Z}$ present in a defected 3D fcc crystal of a given vacancy concentration $c$ by means of a combinatorial argument.
This is done by computing the average connectivity of one lattice site in the vacancy fcc via
\begin{align}
	\bar{Z}
	=
	\sum_{i=0}^{12}i \, \dfrac{{{12}\choose{i}}	{{N-12}\choose{N-N'-i}}	}{ {{N}\choose{N-N'}}}
	=
	12(1- \dfrac{N'}{N})
	 = 12 (1-c).
\end{align}

In terms of the bond occupancy probability $p=\bar{Z}/ 12$ this results in the simple relation  $p=1-c$. 
\begin{figure}[t]
\centering
  \includegraphics[width=\linewidth]{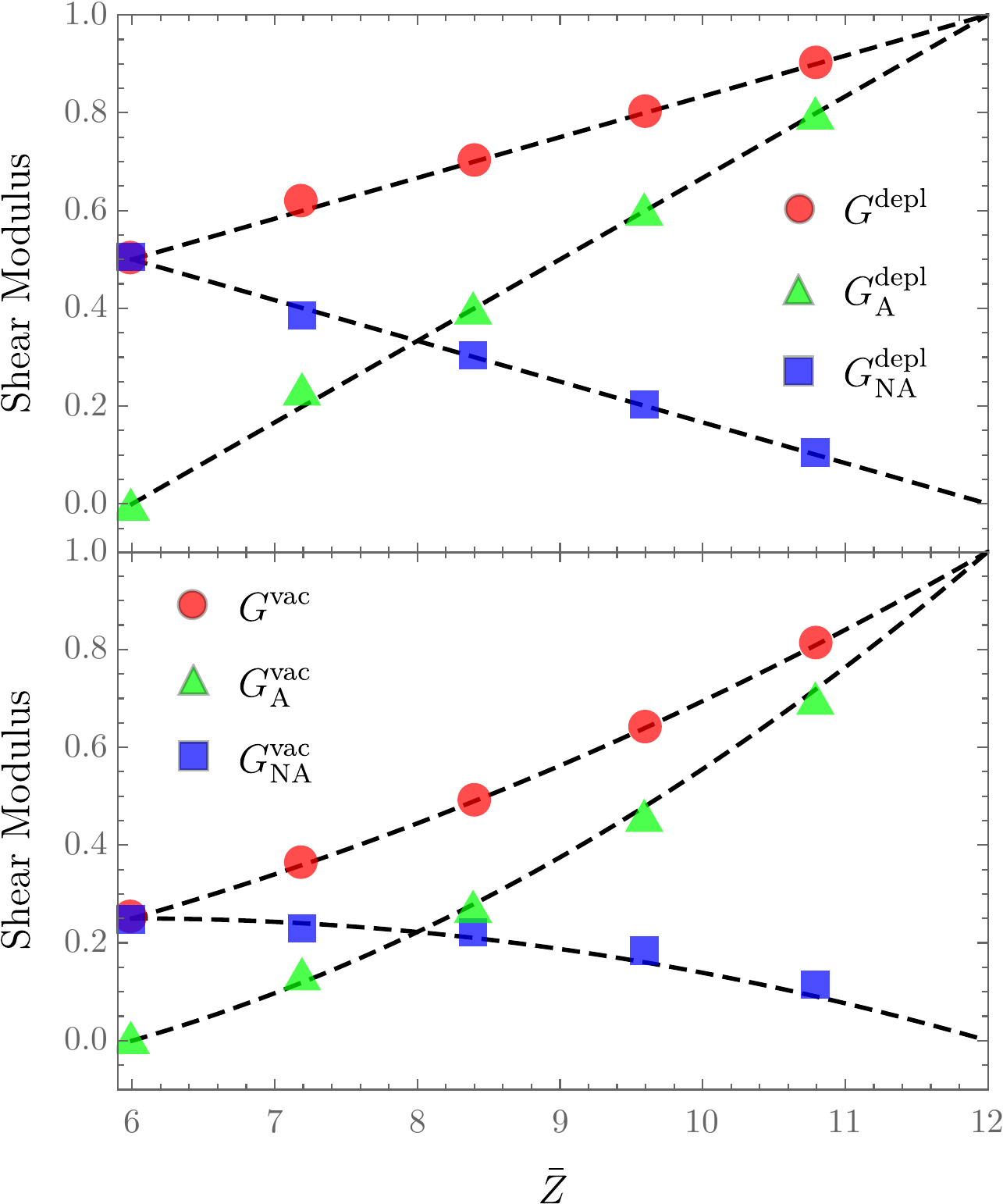}
  \caption{Comparing the shear modulus of the 3D fcc crystal in the case of vacancy defects and bond depletion. The dashed lines are the shear moduli according to Eq.~\ref{eq_shear_depl} in the upper panel and Eq.~\ref{eq_shear_vac} in the lower panel.}
  \label{fig_shear_mod_comp}
\end{figure}
This leads us to ask whether this simple analogy between the depleted and defective fcc carries over to the elasticity and the vibrational properties.
Using an analogous combinatorial argument we can explicitly write down the fluctuation of the $Z$-distribution for the vacancy fcc as
\begin{align}
\sigma_Z^2 &= 
\langle Z ^2\rangle - \langle Z \rangle^2
=
12\dfrac{\left( 1- \frac{12}{N}\right) \left( 1-\frac{N'}{N}\right)\frac{N'}{N}}{1-\frac{1}{N}}
\notag \\&
\xrightarrow{N\to \infty} 12c(1-c)
\label{eq_vac_sigma_Z}
\end{align}
where in the last step $N'/N=c$ is kept constant.

%

Substituting the relation $p=1-c=\bar{Z}/12$ and the resulting new number density $\rho'=\frac{N-N'}{V} = \frac{N}{V} (1-c) = \rho (1-c)$ into to the expression for the bond-depleted fcc shear modulus $G_\text{A}^{\text{depl}}$, we obtain the affine part of the shear modulus with vacancies
\begin{align}
	G_{\text{A}}^{\text{vac}} =\dfrac{\kappa}{2} \rho' r_0^2 \dfrac{\bar{Z}}{12}= \dfrac{\kappa}{a} (1-c)^2.
\end{align}
In the same way, we can transform the non-affine contribution of the bond-depleted fcc shear modulus $G_{\text{NA}}^{\text{depl}}$ and obtain
\begin{align}
G_{\text{NA}}^{\text{vac}} = \dfrac{\kappa}{a}c (1-c),
\end{align}
such that the full vacancy shear modulus becomes
\begin{align}
G^{\text{vac}} 
=&
G_{\text{A}}^{\text{vac}} 
-
G_{\text{NA}}^{\text{vac}} 
=
\dfrac{\kappa}{a} (1-c)(1-2c).
\label{eq_shear_vac}
\end{align}
In terms of the average coordination number $\bar{Z}$ this result reads
\begin{align}
	G^\text{vac} = \dfrac{\kappa}{a} \dfrac{\bar{Z}}{12} \dfrac{\bar{Z}-6}{6}
	=
 \dfrac{\bar{Z}}{12} G^{\text{depl}}.
\end{align}
\subsection{Comparison with simulations}

In order to check the validity of formula~\eqref{eq_shear_vac} derived for the vacancy fcc shear modulus above, we now compare this theoretical prediction to the result of a numerical solution of the non-affine lattice dynamics equations of the vacancy fcc crystal.

The numerical solution of lattice dynamical equation for the shear modulus Eq.~\eqref{eq_G} is based on initiating an perfect fcc crystal of a given density $\rho = N/V$ consisting of 4000 particles interacting via a harmonic potential subject to periodic boundary conditions. Then bonds or particles are removed uniformly at random from the system to reach the desired bond-occupation $p$ or vacancy concentration $c$, respectively. The resulting configuration is then used to solve the equations of motion of non-affine lattice dynamics~\cite{Lemaitre2006} from which the shear modulus can be extracted.

In contrast to the case of a bond-depleted fcc crystal, where we can generate configurations with different distributions peaked around the average connectivity $\bar{Z}$, we have no such control over the $Z$-distribution in the defective crystal with vacancies.

In particular, we generated a bond-depleted fcc crystal with a very narrow $Z$-distribution. In this narrow case the variance of the distribution of the average connectivity $P(Z)$ is $0.07,\,0.23,\,0.24,\,0.24,\,0.16$ for $\bar{Z}=$6, 7.2, 8.4, 9.6, 10.8, respectively.

\begin{figure}[t]
\centering
  \includegraphics[width=\linewidth]{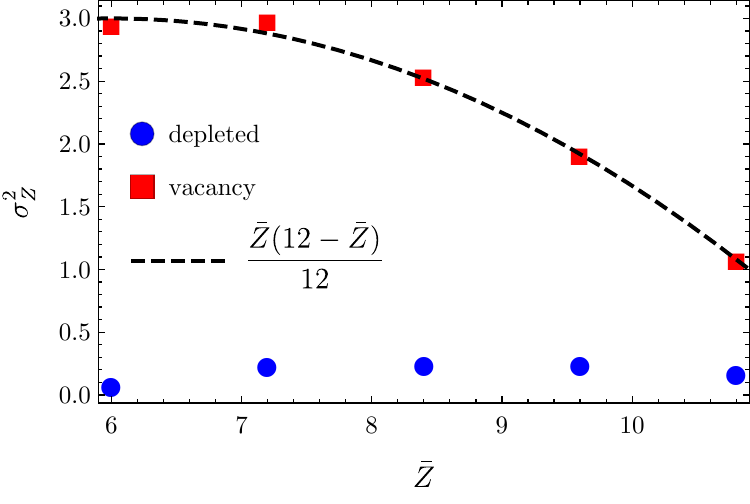}
  \caption{The fluctuations of the ${Z}$-distribution for depleted and vacancy fcc vs $\bar{Z}$.
  This is in accordance with the prediction Eq.~\eqref{eq_vac_sigma_Z} shown as the dashed line.}
  \label{fig_variance_Z}
\end{figure}

The vacancy fcc crystal has a much broader $Z$-distribution. Its variance, which is analytically given by Eq.~\eqref{eq_vac_sigma_Z}, is $2.95,\,2.98,\,2.54,\,1.91,\,1.07$ for $\bar{Z}=$6, 7.2, 8.4, 9.6, 10.8, see also Fig.~\ref{fig_variance_Z}.
The trend of the distribution $P(Z)$ for the vacancy fcc towards smaller variances at smaller vacancy concentrations is due to the saturation of the distribution at $Z=12$, i.e. particles cannot have more than 12 nearest neighbors. This also means that $P(Z)$ is not symmetric around $\bar{Z}$ for low vacancy concentrations.
The numerical solution for the affine and non-affine shear modulus of the vacancy fcc shows perfect agreement with numerical simulations over a broad range of the average coordination number $\bar{Z}$, as depicted in Fig.~\ref{fig_shear_mod_comp}, where we plotted the affine and non-affine contributions separately.


%
%
%
%
%
\section{Vibrational properties in the presence of vacancies}
In this section we turn to the vibrational properties of the vacancy fcc crystal in order to compare it to the well-studied case of the bond-depleted fcc.
%
%
We will see that the low-frequency properties of the vibrational density of states (DOS) are closely related cases when the vacancy concentration is low.
Both the DOS of the vacancy and the bond-depleted fcc were obtained from a direct numerical diagonalization of the underlying Hessian matrix $\underline{\underline{H}}$. 

\begin{figure*}[t]
\centering
  \includegraphics[width=\linewidth]{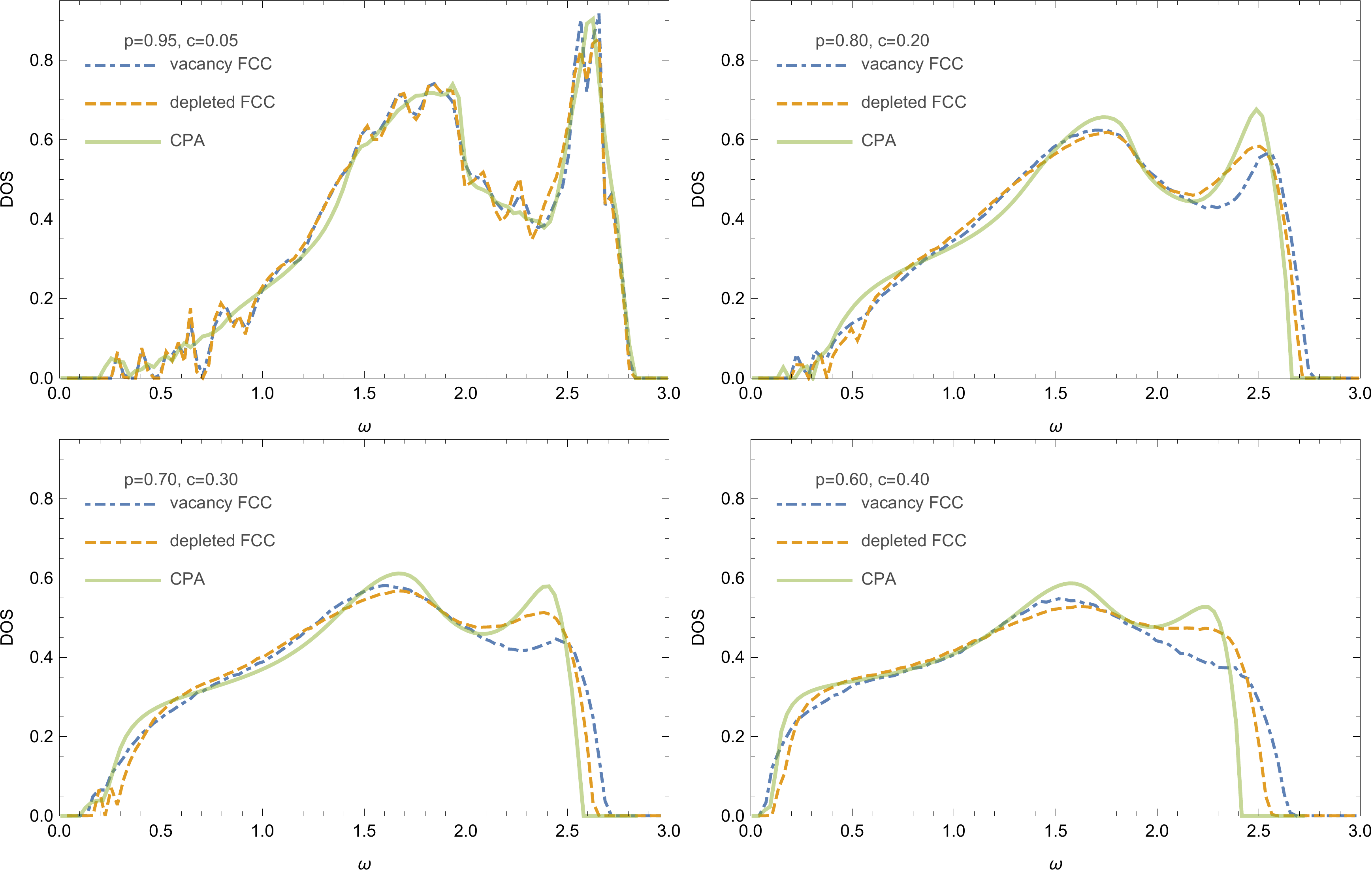}
  \caption{Comparison of the DOS of the depleted fcc with narrow $Z$-distribution and the vacancy fcc to the DOS obtained from the solution of the EMT equations.}
  \label{fig_DOS_comp}
\end{figure*}

In addition to that we computed the DOS of the bond-depleted fcc in the coherent potential approximation\cite{Feng1985,Garboczi1985, Duering2013}, which serves as an effective-medium theory description for the fcc with randomly cut bonds, characterized by the bond-occupation probability $p=\bar{Z}/12$. 
There is very good agreement between the vacancy fcc, depleted fcc and the EMT solution across the whole frequency range for a low vacancy concentration. This is illustrated in Fig.~\ref{fig_DOS_comp} for a vacancy concentration $c=0.05$.
Increasing $c$, or alternatively $p$, the agreement becomes less reliable. In fact, the DOS obtained with CPA does not precisely capture the form of the DOS of the bond-depleted fcc, despite the fact that the bond-depletion was carried out such that the $Z$-distribution is very narrow to resemble the effective-medium solution, which does not account for fluctuations in the connectivity.
Partly this happens due to the fact that CPA does not properly account for scattering from pairs of defects. The effective-medium description only is accurate when the separation between defects is large enough such that the amplitude of a scattered wave is negligible at the neighboring defect~\cite{Crawford1975} .

\subsection{Boson peak scaling}
A universal feature of the vibrational density of states $D(\omega)$ of disordered solids and glasses is the excess of low-frequency modes with respect to the Debye scaling $D(\omega)\sim \omega^2$, which manifests itself as a peak in the vibrational density of states, widely referred to as the boson peak~\cite{Binder2011}.

\begin{figure}[t]
\centering
  \includegraphics[width=\linewidth]{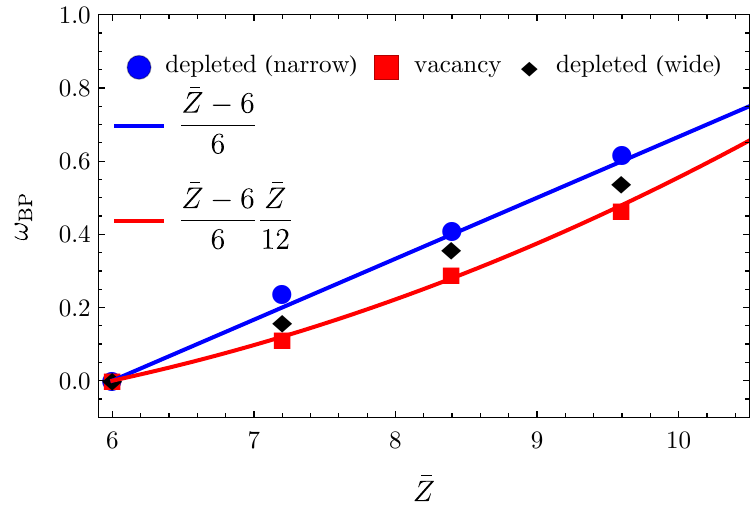}
  \caption{Scaling of the boson peak frequency $\omega_{\text{BP}}$ with $\bar{Z}$ for the three model fcc crystals with different disorder.
  The value of $\omega_\text{BP}$ for the largest coordination number $\bar{Z}=10.8$} is not included in the plot since the boson peak is not identifiable in the DOS. 
  \label{fig_omega_BP}
\end{figure}
Since the appearance of the boson peak is a feature inherently rooted in the disorder in a physical system, the  position of the boson peak in the vibrational spectrum (the boson peak frequency $\omega_\text{BP}$) in disordered fcc crystals depends on the average coordination number $\bar{Z}$, which sets the magnitude of disorder.
In particular, in the case of the bond-depleted fcc (and the random spring network) it is known to have the scaling $\omega_{\text{BP}} = \frac{(\bar{Z} - 6)}{6}$.

This numerically computed scaling of $\omega_\text{BP}$ for the bond-depleted fcc is plotted in Fig.~\ref{fig_omega_BP} and, interestingly, the same scaling does not hold true for the fcc crystal with vacancies. In fact, the boson peak frequency $\omega_{\text{BP}}$ of the vacancy fcc features a parabolic scaling with the average connectivity $\bar{Z}$, which is fitted well by $\omega_{\text{BP}} = \frac{ \bar{Z}-6}{6} \frac{\bar{Z}}{12}$.
It appears that for both the bond-depleted and vacancy fcc, the boson peak frequency exhibits the same scaling with $\bar{Z}$ as the respective shear modulus $G$ with $\bar{Z}$, see Tab.~\ref{tab_1}.

This, however, is not generally the case.  To see this we generated a depleted fcc with a $Z$-distribution which is not $\delta$-like, but resembles $P(Z)$ in the vacancy fcc, i.e. a certain degree of connectivity fluctuations is introduced.
The shear modulus scaling in this wide-depleted case is still the same as in the $\delta$-like depleted case, i.e. $G^{\text{wide,depl}} \sim \frac{\bar{Z}-6}{6}$.
However, the scaling of the boson peak frequency $\omega_\text{BP}$ now is different, as can be seen in Fig.~\ref{fig_omega_BP}.
This rules out the possibility that $G$ and $\omega_{BP}$ generally scale in the same way in our defective fcc systems.

Taking all this into account, the transition between the scalings of $\omega_\text{BP}^\text{depl}$ and $\omega_\text{BP}^\text{vac}$ is not given by the density transformation $\rho' = \frac{\bar{Z}}{12} \rho$.
In fact, we assert that the boson peak is not primarily controlled by the average coordination number $\bar{Z}$, as is the case for the shear modulus. Fluctuations of the Z-distribution will be of importance here, i.e. the degree of heterogeneity of the underlying microstructure will influence the scaling of the boson peak with $\bar{Z}$.
Directly comparing the two different depleted fcc models, we see that increasing the connectivity fluctuations pushes the boson peak frequency to lower values. We will shed further light on this in the next section.

\section{The inversion symmetry breaking parameter}

We observed that for both the bond-depleted and vacancy fcc crystal the scalings of the shear modulus with the average connectivity $\bar{Z}$ coincide, whereas the same cannot be concluded for the scaling of $\omega_\text{BP}$ for the two types of systems.
This effect must in some way be connected to the microstructural differences which arise through the different implementations of disorder in the bond-depletion and vacancy situation.
With the aim of bringing a physical justification to the two different scalings of the shear modulus and the boson peak frequency we need to quantitatively describe the degree of local microstructural disorder. 

%
%
%
%
The starting point is defining a parameter which serves as a measure for the degree of local inversion symmetry, by which we mean the condition that each nearest neighbor of a reference particle has a mirror particle diametrically opposed.
The squared amplitude of the affine force vector $\vert \underline{\Xi}_i\vert^2$ precisely serves this purpose, measuring the local deviations from the case of perfect inversion symmetry. It is identically zero in a centrosymmetric crystal, where in the affine configuration, the local structure of nearest neighbors around a particle is such that the positions of two opposing neighbors are symmetric with respect to a reflection at the central particle.
In this case the square of the affine force vector $\vert \underline{\Xi}\vert^2$  is exactly zero, as mentioned above.
For the other limiting case where the system completely lacks local inversion symmetry, we chose a reference configuration to normalise the ISB parameter with the corresponding squared affine force field. For this reference configuration we require that there be no correlations between the orientations of the bonds~\cite{Milkus2016}.
In this way we obtain a measure for the local inversion symmetry which varies between zero and one.

The ISB parameter derived from $\vert \underline{\Xi}\vert^2$ should be independent of the direction of the applied shear stress, which means it has to be summed over all possible coordinate pairs as
\begin{align}
	\vert \underline{\Xi}\vert^2
	:=
	\sum_{\alpha, \beta \in [x,y,z]} 
	\vert \underline{\Xi}_{\alpha \beta}\vert^2.
\end{align}
Following the exposition in Ref.~\cite{Milkus2016}, we define the parameter for measuring the local inversion symmetry breaking as
\begin{align}
	F_{\text{IS}}
	=&
 1 - \dfrac{\langle \vert \underline{\Xi}_i\vert^2 \rangle}{\langle \vert \underline{\Xi}^{\text{RI}}_i\vert^2 \rangle}
	\notag \\=&
	1-
	\dfrac{\sum_{\alpha, \beta \in [x,y,z]} \vert \underline{\Xi}_{\alpha \beta}\vert^2}{\sum_{\alpha, \beta \in [x,y,z]} \vert \underline{\Xi}_{\alpha \beta}\vert^2_{\text{ISB}}}
	\notag \\=& 
	-\dfrac{1}{N} \sum_i^N \dfrac{1}{Z_i} \sum_{j,k \;n.\,n. \;i } \cos^3\alpha_{jk}
\end{align}
where $\alpha_{jk}$ denotes the angle between the $i$-$j$ and $i$-$k$ bonds.
A more detailed derivation of the above expression can be found in the Appendix. 
We now proceed to evaluate the degree of local inversion symmetry breaking in the two fcc crystals with bond-depletion and vacancy induced disorder.


\subsection{Inversion symmetry breaking in the depleted fcc}
\begin{figure}[t]
\centering
  \includegraphics[width=\linewidth]{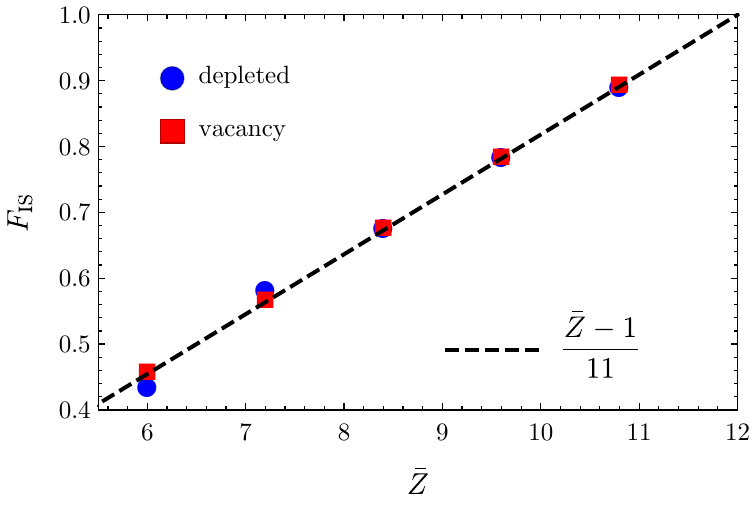}
  \caption{Comparison between the average ISB parameter for the vacancy and depleted fcc}
  \label{fig_FIS_comp}
\end{figure}

As it was demonstrated in earlier work~\cite{Milkus2016}, it is possible to derive a analytical expression for the ISB parameter in the case of the bond-depleted fcc, with the result~\cite{Milkus2016}
\begin{align}
F_{\text{IS}}^{\text{depl}} = 1- \dfrac{\sum_{\alpha, \beta}	\vert \underline{\Xi}_{\alpha \beta} \vert	}{R_0^2 \kappa^2 N \bar{Z} }
		= 1 - \dfrac{12-\bar{Z}}{11}
		= \dfrac{\bar{Z}-1}{11}.
\end{align} 
This perfectly aligns with the numerical evaluation of $F_{\text{IS}}^{\text{depl}}$, as can be seen from Fig.~\ref{fig_FIS_comp}.  From there we can also observe that the numerical computation of the inversion symmetry breaking parameter $F_{\text{IS}}^{\text{vac}}$ for the vacancy fcc yields virtually the same linear behavior as in the depleted case. From a physical point of view, we can say that the average 
ISB parameter does not distinguish between vacancy and depleted fcc. This is because the ISB is defined in terms of the angles between NN particles in a unit cell. Hence, the average ISB being equal in the vacancy and depleted case means that the distribution of angles between next neighbors is the same on average. This is reasonable because particles or bonds are removed at uniformly at random in both cases.

Naturally, this leads us to conjecture that the functional form in both cases is the same, i.e.
$
	F_{\text{IS}}^{\text{depl}}=F_{\text{IS}}^{\text{vac}} = \frac{1}{11} (\bar{Z}-1).
$
We have, however, not yet been able to verify this result analytically for the vacancy case.
%
%
%
The physical picture behind this observation is as follows. The two different approaches of inducing disorder into the fcc crystal, i.e. by removing bonds or particles, produce a disordered microstructure in the crystal which on average exhibits the same degree of local inversion symmetry breaking.
In fact, we can use this argument to justify why the shear moduli of the depleted and vacancy case differ only by the density correction due to the missing particles in the defected case. 

When computing the elastic constants, only the average degree of disorder of the microstructure or, put differently, only the averaged degree of inversion symmetry breaking controls the shear modulus. Just as the ISB parameter, the non-affine contribution to the shear modulus is proportional to the averaged squared amplitude of the affine force field $\underline{\Xi}$, i.e.
$	G_{\text{NA}} \propto \langle \vert 	\underline{\Xi}_i	\vert^2\rangle
$~\cite{Lemaitre2006, Milkus2016}.

More specifically, for the computation of the zero-frequency shear modulus in the thermodynamic limit one needs to solve the integral~\cite{Lemaitre2006}
\begin{align}
G(\omega = 0) = G_\text{A} - \dfrac{3N}{V}\int_0^\infty \mathrm{d} \omega'\dfrac{\rho (\omega') \Gamma(\omega' )}{m \omega'^2}
\label{G_integral}
\end{align}
where $\rho(\omega')$ is the density of states and $\Gamma(\omega')$ the correlator of the affine force fields between frequency shells. This correlator is defined as
\begin{align}
	\Gamma(\omega) = \big \langle ( \underline{\Xi}\cdot \underline{v}_p)^2	 \big \rangle
\end{align}
where $\underline{v}_p$ is the eigenvector of the Hessian $\underline{\underline{H}}$, which belongs to the eigenfrequency $\omega_p$. The average is performed for all projections of the affine force fields onto eigenvectors with eigenfrequency $\omega_p \in [\omega, \omega + \mathrm{d}\omega]$.
Since the density of states appears together with the correlator under the integral, similar features in the density of states are not sufficient to guarantee the same behavior in terms of the elastic moduli. In this sense, the shear modulus is a coarse-grained, macroscopic physical quantity.

The physical mechanism responsible for the loss of mechanical stability, which fundamentally is based on the concept of local inversion symmetry is the same for both the depleted and vacancy fcc. We have seen above that the density enters as a prefactor into the formula of the shear modulus. By the above argument, the quantities inside the parentheses of \eqref{eq_G} are the same for the bond-depleted and vacancy case, which leaves the corrected density as the only source of the different scaling of the shear modulus of the vacancy fcc.

\subsection{Correlation of the boson peak with ISB fluctuations}

As we have seen, the positions of the boson peak of the depleted and vacancy fcc generally show different correlations with the average connectivity $\bar{Z}$. But unlike in the case of the shear modulus scaling, the boson peak position cannot only depend on the density correction due to vacancies.
This also indicates that the behavior of $\omega_\text{BP}$ is not exclusively dictated by the average degree of inversion symmetry breaking. The shear modulus in contrast, being a macroscopic averaged quantity, is not sensitive to the fluctuations of the distributions $P(Z)$ and $P(F_\text{IS})$ for the disordered fcc crystals.

\begin{figure}[t]
\centering
  \includegraphics[width=\linewidth]{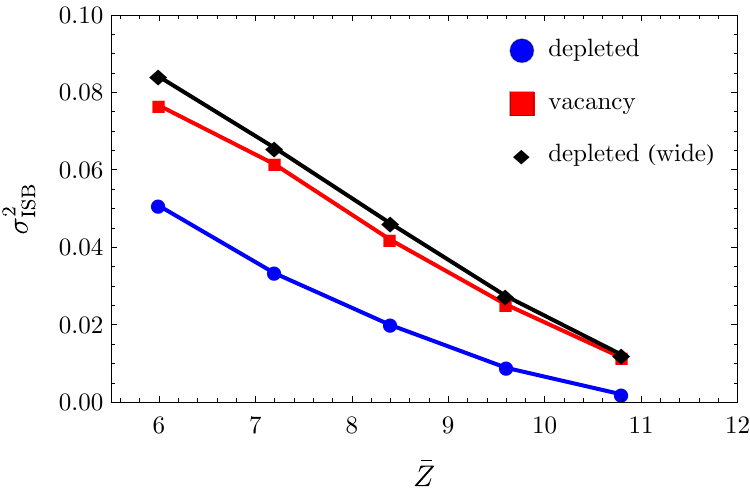}
  \caption{Plot of the $\bar{Z}$-dependence of the variance of the distribution $P(F_{\text{IS}})$.}
  \label{fig_variance_FIS}
\end{figure}

To better understand the origin of the boson peak scaling we need to take these fluctuations into account.
We numerically computed the distributions of the connectivity and ISB parameter. The resulting widths of these distributions are plotted in Fig.~\ref{fig_variance_Z} and Fig.~\ref{fig_variance_FIS}. Naturally, the fluctuations of $Z$, i.e. the structural heterogeneity of the disordered fcc crystal, is directly linked to the fluctuations of $F_\text{IS}$. The $\delta$-like distribution of the coordination number in the depleted fcc is linked to a microstructure which is less heterogenous when compared to the vacancy fcc. 

Reducing the average connectivity towards the mechanical instability at $Z_c$, the fluctuations of the ISB monotonically increase. But for any given value of $\bar{Z}$, the fluctuations of the ISB are larger in the vacancy fcc. This behavior is reflected in the fact that $P(F_\text{IS})$ is a broader distribution in the vacancy case. This relative broadness goes hand in hand with a more asymmetric distribution: the vacancy $P(F_\text{IS})$ has an excess at low values of the ISB parameter with respect to the depleted crystal case. Physically speaking, this means the vacancy fcc has an excess of sites with low inversion symmetry compared to the $\delta$-like depleted fcc.
So the higher fluctuations of $F_\text{IS}$ in the vacancy fcc go along with a higher degree of asymmetry of $P(F_\text{IS})$. As a consequence, the vacancy fcc develops highly undercoordinated sites much earlier than the bond-depleted fcc, when decreasing $\bar{Z}$.
We can conclude that for the defective fcc crystals studied here that in the above sense structural heterogeneity modifies the boson peak via increased ISB fluctuations, such that $\omega_{BP}$ is pushed to lower frequencies.
We have collected the scalings of the shear modulus and the ISB with $\bar{Z}$ in Tab. 1 for both the depleted and vacancy fcc.

In recent work\textcolor{red}{~\cite{Ganguly2013,Ganguly2015}} a similar connection has been observed in a system where non-affine displacements are induced via thermal fluctuations instead of structural disorder, which complements our findings. The analytical and numerical results\textcolor{red}{~\cite{Ganguly2017}} show that if the non-affine displacements are enhanced via an external field there is an accumulation of low-frequency modes in the density of states, also linking non-affine fluctuations to the boson peak.

\section{Application to colloidal crystals}
\begin{figure}[h]
\centering
  \includegraphics[width=\linewidth]{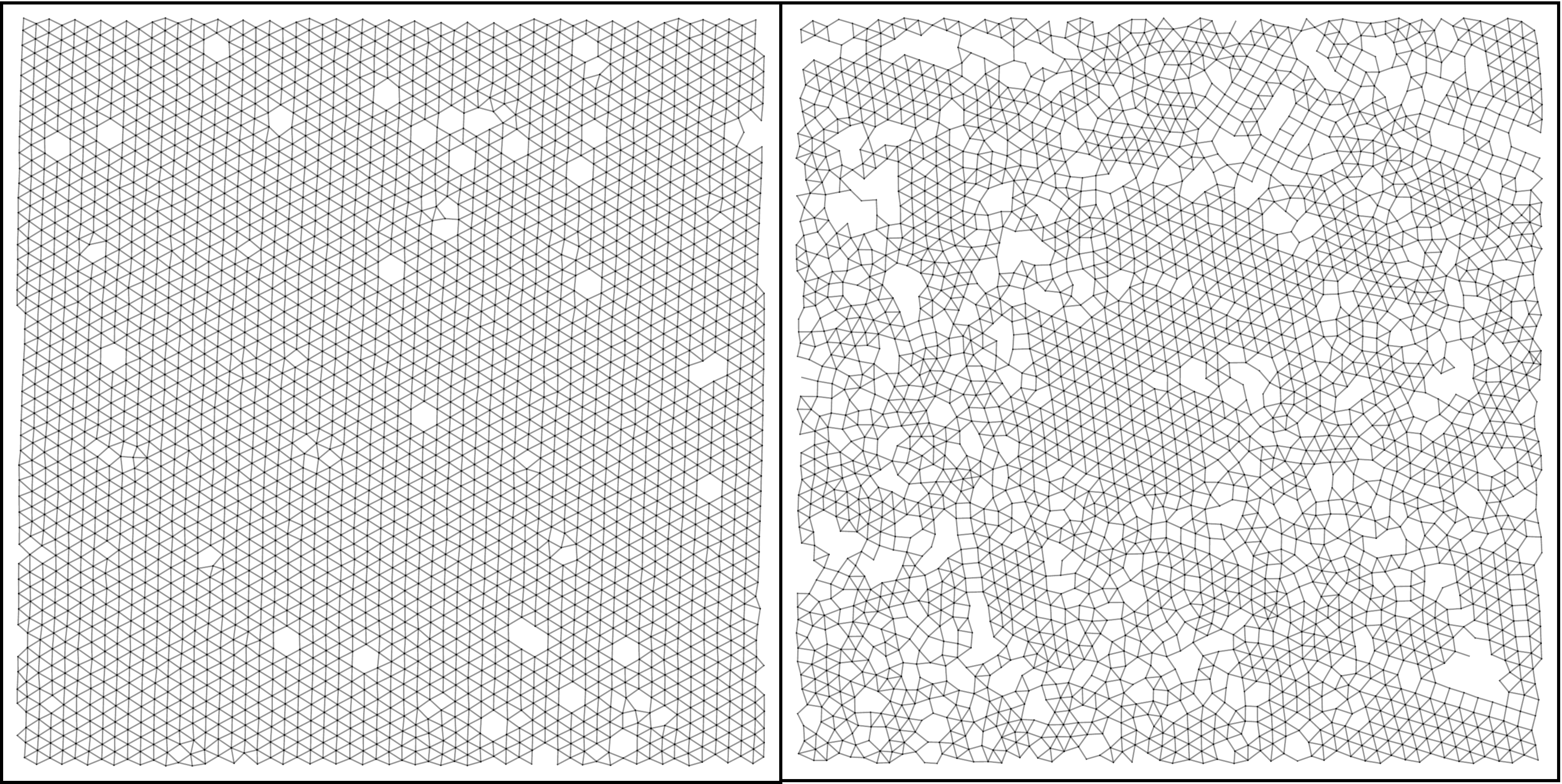}
  \caption{Comparison of 2D slices of the vacancy fcc colloidal crystal prepared in Ref.~\cite{Zargar2014} for low ($c=0.0167$) and high ($c=0.169$) vacancy concentration both at a volume fraction $\phi=0.56$.}
  \label{fig_lattice_vac_comp}
\end{figure}

\begin{figure}[t]
\centering
  \includegraphics[width=\linewidth]{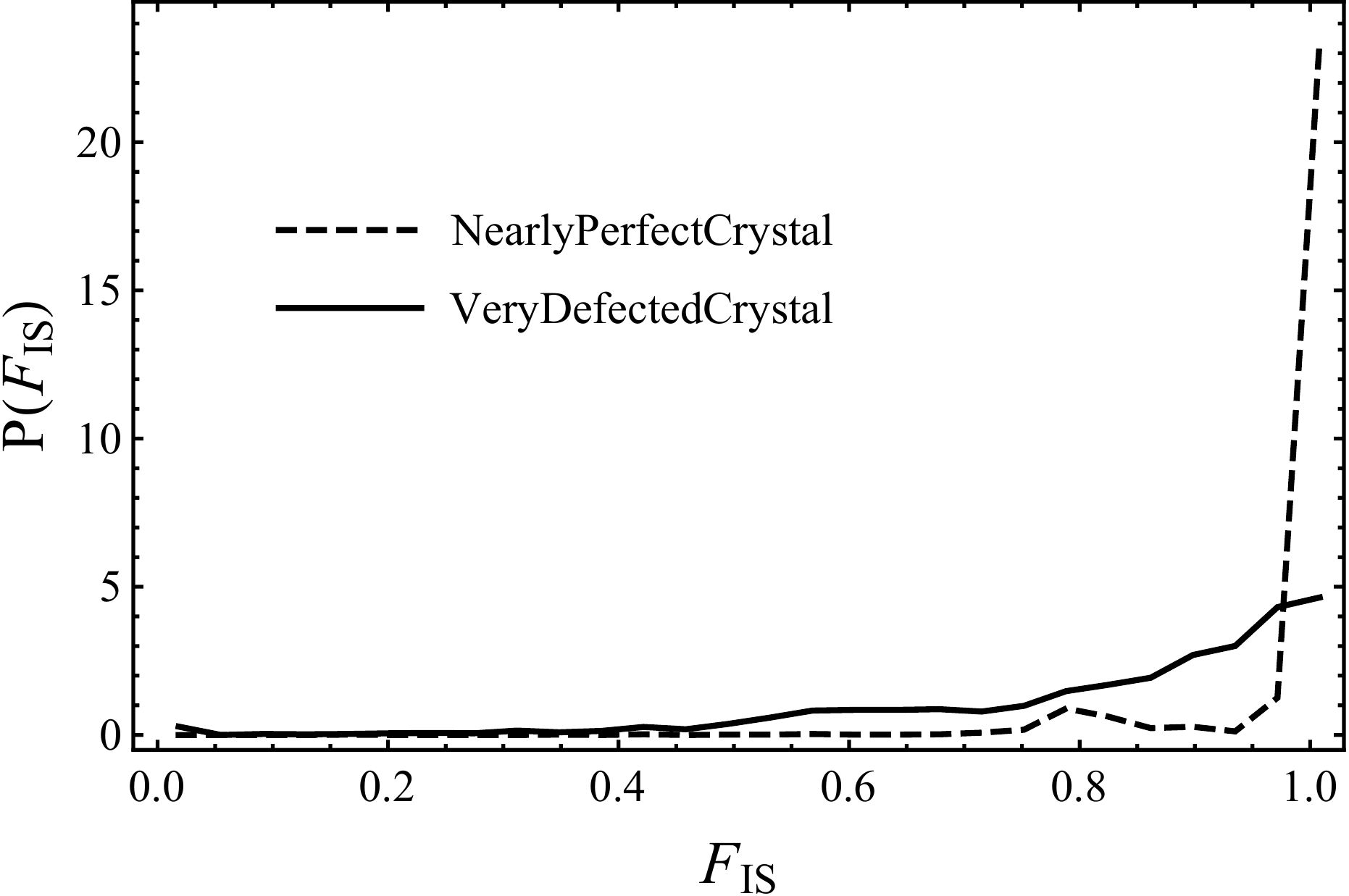}
  \caption{The inversion symmetry parameter $F_\text{IS}$ computed from the two 2D low and high vacancy concentration configurations shown in Fig.~\ref{fig_lattice_vac_comp} }
  \label{fig_FIS}
\end{figure}
Using the experimental input from Zargar et al.~\cite{Zargar2014} we can obtain a numerical estimation for the absolute value of the shear modulus of a defective colloidal fcc crystals with vacancies.
	Two examples of configurations with a low ($c=0.0167$) and high ($c=0.169$) vacancy concentration are shown in Fig.~\ref{fig_lattice_vac_comp}.
In these systems the defect concentration of a sample is determined using confocal microscopy.

To fix a numerical value of the shear modulus $G^{\text{vac}}(c)$ we also need the values  of the lattice constant $a$ of the colloidal crystal and its spring constant. In the experimental setup of Zargar et al., $a$ is given by $3.019 \mu m$. The spring constant has to be estimated from the confocal microscopy data by determining the potential of mean force from the measured radial distribution function of the colloidal system in the liquid phase. The resulting main minimum of the potential of mean force is subsequently fitted with a harmonic potential, which in the present case fixes the spring constant to $\kappa = 9.24 \cdot 10^{-7} N/m$. 
Together with Eq.~\eqref{eq_shear_vac}, we thus can estimate the vacancy shear modulus to be
$G^{\text{vac}}(c=0) = 0.306 \;\text{Pa}$, $G^{\text{vac}}(c=0.0167) = 0.291 \;\text{Pa}$ and $G^{\text{vac}}(c=0.169) = 0.168 \;\text{Pa}$.
Thus, Eq.~\eqref{eq_shear_vac} predicts that the shear modulus of the crystal with vacancies should decrease by 50$\%$ when about 17$\%$ of the available lattice site are vacant.

\section{Conclusions}

Using the non-affine lattice dynamical expressions and combinatorial bond counting, we found that 
the expressions for the shear moduli of bond-depleted and vacancy fcc crystals share an equivalent formal structure. They are connected via the transformation $p = 1-c$ and a density rescaling proportional to $1-c$, (where $p$ is the fraction of depleted bonds and $c$ the concentration of vacancies). This behavior can be attributed to the fact that the average or global degree of inversion symmetry breaking controls the shear modulus of the disordered fcc crystal, and this inversion-symmetry breaking (ISB) parameter exhibits the exact same behavior in the two systems under investigation (i.e. bond-depleted and vacancy lattice, respectively).

When describing the elastic properties of disordered solids, the equations of non-affine lattice dynamics reflect the fact that besides the standard Born-Huang affine contribution there is an additional non-affine contribution which leads to an effective elastic softening.
This softening mechanism is closely related to the affine force field $\underline{\Xi}_i$ which represents the additional forces acting on a particle $i$ due to the disorder-induced imbalance of forces, an effect rooted in the absence of local inversion symmetry in a disordered lattice.

In order to compute the non-affine correction to the shear modulus we have to average the affine force fields with respect to the disordered configurations of the crystal. So the shear modulus cannot depend on the local properties of the local degree of inversion symmetry breaking, because they are washed out by self-averaging in the process of taking the thermodynamic limit. 
 
Furthermore, comparing the vibrational properties of the two disordered fcc crystals, we saw that the density of states of both systems are closely related up to high degrees of disorder. However, we also observed that the boson peak position $\omega_\text{BP}$ is generally shifted to lower frequencies in the case of the vacancy fcc. This is because the fluctuations of the ISB parameter, more than its mean value, control the behavior of the boson peak.

There are two closely linked quantities which account for the microstructure of the disordered fcc.
The distribution of the connectivity $Z$ can be intuitively understood.  A narrowly peaked distribution, which was used for the bond-depleted fcc in this work, means that every particles in the system has approximately the same number of nearest neighbors with a high probability.
On the other hand, a broad $Z$-distribution tells us that the heterogeneity of the microstructure, and hence of the variation in the number of nearest neighbors, is large. 

The connection to the inversion symmetry breaking parameter is made easily.  A configuration with a large degree of $Z$ spatial fluctuation directly leads to a heterogenous distribution of the ISB parameter. Highly coordinated sites retain most of their bond symmetry and lead to $F_\text{IS}$ values close to one. A larger spread of local coordination numbers $Z$ yields a broader distribution $P(F_\text{IS})$.
When the $Z$-fluctuations in the system are large we can conclude that we have a relative excess of
low-coordinated particles with respect to a system with a narrower $Z$-distribution.
Since these low-coordinated sites are connected to a smaller number of bonds, it takes a smaller amount of energy to excite vibrations at this site due to the fact that they effectively have a lower binding energy.

In turn, these low-energy excitations translate to an increased population of modes at low frequencies. Therefore, the low-frequency part of the DOS, together with the boson peak at $\omega_\text{BP}$, moves to lower frequencies.

Following this argument, the boson peak of the vacancy fcc is always shifted to lower frequencies with respect to the depleted fcc due to larger fluctuations of the local degree of inversion symmetry.
This behavior is clearly seen from the numerical solution of the non-affine lattice dynamics of the depleted and vacancy fcc, which also holds true for the depleted fcc with a wide $Z$-distribution.

In conclusion, our analysis of the vibrational properties and shear elasticity within the framework of non-affine lattice dynamics, has made it possible to identify the microscopic source and the nature of the "disorder" and heterogeneous fluctuations that have served in previous theoretical studies based on fluctuating elasticity models~\cite{Maruzzo2013,Schirmacher2006,Schirmacher2015} as the input to explain the boson peak in disordered solids and its link with soft elasticity.

\begin{table}[h]
\small
  \caption{\ Scaling of different physical quantities with $Z$ for the depleted and vacancy fcc.}
  \label{tab_1}
  \begin{tabular*}{0.48\textwidth}{@{\extracolsep{\fill}}lll}
    \hline
    Quantity & Depleted & Vacancies \\
    \hline
    \vspace{2mm}
    $F_{\text{IS}}\vphantom{\int}$& $\dfrac{\bar{Z}-1}{11}$& $\dfrac{\bar{Z}-1}{11}$ \\
      \vspace{2mm}
      \vspace{2mm}
    $G_{\text{A}}$ & $\dfrac{\bar{Z}}{12}$ & $\dfrac{\bar{Z}^2}{12^2}$ \\
      \vspace{2mm}
    $G_\text{NA}$ & $\dfrac{12-\bar{Z}}{12}$ & $\dfrac{\bar{Z}}{12} \dfrac{12-\bar{Z}}{12}$ \\
     \vspace{2mm}
    $G $& $\dfrac{\bar{Z}-6}{6}$ & $\dfrac{\bar{Z}}{12}\dfrac{\bar{Z}-6}{6}$ \\
    \vspace{2mm}
    $\omega_{\text{BP}} $& $\dfrac{\bar{Z}-6}{6}$ & $\dfrac{\bar{Z}}{12}\dfrac{\bar{Z}-6}{6}$ \\
    \hline
  \end{tabular*}
\end{table}

\newpage
\section{Appendix}
\subsection{EMT for simple spring networks}

\newcommand{\bra}[2][]{\mathinner{\langle #2\rvert}_{#1}}
		\newcommand{\ket}[2][]{\mathinner{\lvert#2\rangle}_{\hspace{-0.1em}#1}}

We start with the $N$-particle Hamiltonian in $d$ dimensions (e.g. of the triangular lattice)
\begin{align}
\mathcal{H}= \mathcal{H}_0 + \mathcal{V}
\intertext{where}
\mathcal{H}_0 = \mathcal{H}_{\underline{p}}+ \mathcal{H}_{\underline{u}}
\end{align}
where $\underline{p}$ and $\underline{u}$ are the $Nd$-dimensional momentum and displacement vector, respectively.
For ease of notation, we write the $\underline{u}$ as
\begin{align}
\ket{\underline{u}} = \left(	\underline{u}_1, \dots , \underline{u}_N	 \right)
\intertext{with the property}
\langle i \rvert \underline{u} \rangle = \underline{u}_i.
\end{align}
In the harmonic approximation $\mathcal{H}_{\underline{u}}$ is given by
\begin{align}
\mathcal{H}_{\underline{u}} = \dfrac{1}{2} 
	\sum_{\langle i,j \rangle}
	k_{ij}
	\left[ \vphantom{\sum}	(\underline{u}_i - \underline{u}_j)\cdot \hat{\underline{r}}_{ij}	\right]^2
\end{align}
where ${\langle i,j \rangle}$ means summing over next neighbors and $\hat{\underline{r}}_{ij}$ are the unit bond vectors.
Introducing the $dN \times dN$ dynamical matrix $\mathcal{M}$ we may write
\begin{align}
\mathcal{H}_{\underline{u}} & =  \bra{\underline{u}} \mathcal{M} \ket{\underline{u}}
\intertext{with}
\mathcal{M}&= \dfrac{1}{2}\sum_{\langle i,j \rangle}k_{ij} \hat{\underline{r}}_{ij}\hat{\underline{r}}_{ij}^T \left( \ket{i} - \ket{j}\right) \left( \bra{i} - \bra{j}\right)
\\
&=
\sum_{\langle i,j \rangle} k_{ij} \hat{\underline{r}}_{ij}\hat{\underline{r}}_{ij}^T \mathcal{P}_{ij}
\end{align}
and define for later use the bond projector $\mathcal{P}_{ij}=\frac{1}{2}\left( \ket{i} - \ket{j}\right) \left( \bra{i} - \bra{j}\right)$ \cite{Kirkpatrick1973}.
Accordingly we define the Green's function of the system as
\begin{align}
\mathcal{G}(\omega) = \left[ \mathcal{M} - m \omega^2\right]^{-1}
\end{align}
Following~\cite{DeGiuli2014,Duering2013}, one way to introduce EMT and compute the disorder-averaged Green's functions is to introduce an effective spring constant $k_{\text{eff}}$ and write $k_{ij} =k_{\text{eff}}+
( k_{ij}-k_{\text{eff}})$ and, correspondingly, decompose the dynamical matrix as $\mathcal{M} = \mathcal{M}_0 + \delta \mathcal{M}$ such that 
\begin{align}
\mathcal{M}_0 = \sum_{\langle i,j \rangle} k_{\text{eff}} \hat{\underline{r}}_{ij}\hat{\underline{r}}_{ij}^T \mathcal{P}_{ij}
\intertext{and}
\delta\mathcal{M} = \sum_{\langle i,j \rangle} ( k_{ij}-k_{\text{eff}}) \hat{\underline{r}}_{ij}\hat{\underline{r}}_{ij}^T \mathcal{P}_{ij}.
\end{align}
We then can write the Green's function as $\mathcal{G}=\mathcal{G}_0+\mathcal{G}_0 \mathcal{T}\mathcal{G}_0$, where $\mathcal{G}_0=\left[ \mathcal{M}_0 - m \omega^2\right]^{-1}$ is the Green's function of the effective medium. The matrix $\mathcal{T}$ is the scattering matrix and given by $\mathcal{T}= \delta\mathcal{M}\left[1- \mathcal{G}_0\delta \mathcal{M} \right]^{-1}$, or equivalently,
\begin{align}
\mathcal{T} &= \delta \mathcal{M} + \delta \mathcal{M}\mathcal{G}_0\delta \mathcal{M}+ \delta \mathcal{M}\mathcal{G}_0
 \delta \mathcal{M}\mathcal{G}_0\delta \mathcal{M}+\dots
 \\ &=
\delta \mathcal{M} \sum_{n=1}^\infty \left[ \vphantom{\sum} \mathcal{G}_0\delta\mathcal{M}  \right]^n.
 \end{align}
Inserting the expression for $\delta \mathcal{M}$ into the above series expansion, we can split the sums over the NN-bonds into diagonal and off-diagonal parts. Since $\mathcal{P}_{ij}$ is a projection operator the contributions containing different powers of the projection can be collected and resummed. The result of this is
\begin{align}
	\mathcal{T} =  \sum_{\langle i,j \rangle} \underline{T}_{ij}
	+
	\sum_{\langle i,j \rangle \neq \langle m,n \rangle} \underline{T}_{ij} \mathcal{G}_0 \underline{T}_{mn} +\dots
\end{align}
with
\begin{align}
\underline{T}_{ij} = \dfrac{\left( \ket{i} - \ket{j} \right)\big( k_{ij} - k_{\text{eff}}\big)\left( \bra{i} - \bra{j} \right)}
{1-\left( \vphantom{\sum} k_{ij} - k_{\text{eff}}\right)
\hat{\underline{r}}_{ij}^T	\left( \bra{i} - \bra{j} \right)\mathcal{G}_0\left( \ket{i} - \ket{j} \right)	\hat{\underline{r}}_{ij}
}
\hat{\underline{r}}_{ij}\hat{\underline{r}}_{ij}^T
\end{align}
We now want to determine the effective spring constant $ k_{\text{eff}}$ such that it mimics the average behavior of the disordered system~\cite{Duering2013}, i.e. $\langle \mathcal{G} \rangle = \mathcal{G}_0$. This leads to the condition $\langle \mathcal{T}\rangle =0$, which in the EMT is achieved by setting $ \langle \underline{T}_{ij}\rangle =0$. We evaluated the average over the disorder according to
\begin{align}
0=p \,\underline{T}_{ij}\big\rvert_{ k_{ij}=k} +(1-p) \,\underline{T}_{ij} \big\rvert_{ k_{ij}=0}
\end{align}
meaning that bond randomly removed with probability $1-p$. After some manipulations we can rewrite the above as
\begin{align}
\underline{\hat{r}}_{ij}^T \cdot \left(\vphantom{\sum}\left( \bra{i} - \bra{j} \right)\mathcal{G}_0\left( \ket{i} - \ket{j} \right)	 \right)  \cdot \underline{\hat{r}}_{ij}
=
\dfrac{k_{\text{eff}}-pk}{k_{\text{eff}}(k-k_{\text{eff}})}.
\end{align}
The un-disordered, effective lattice is isotropic and homogeneous so that the above equation is independent of  the bond label $i,j$ \cite{Kirkpatrick1973,DeGiuli2014}. Hence, $\bra{i} \mathcal{G}_0 \ket{i}=\bra{j} \mathcal{G}_0 \ket{j}$ and $\bra{j} \mathcal{G}_0 \ket{i} = \bra{i} \mathcal{G}_0 \ket{j}$.
Using the relation $\mathcal{G}_0(\mathcal{M}_0- m \omega^2) = 1$ and computing its trace we find
\begin{align}
1 + \dfrac{m \omega^2}{d} \text{Tr} \mathcal{G}_0 = \dfrac{k_{\text{eff}}z}{d}\underline{\hat{r}}_{ij}^T \cdot \left(\vphantom{\sum}\left( \bra{i} - \bra{j} \right)\mathcal{G}_0\left( \ket{i} - \ket{j} \right)	 \right)  \cdot \underline{\hat{r}}_{ij}
\end{align}
which corresponds to the final result obtained by Feng and Thorpe \cite{Feng1985}, except for the sign convention of the Green's function. Following \cite{Garboczi1985} we set $\text{Tr}  \mathcal{G}_0 = G_{11}$, giving the magnitude of the site-diagonal Green's function. We can then combine the last two equations to obtain a simplified expression for $G_{11}$ as
\begin{align}
m \omega^2 G_{11} + 1 
= \dfrac{k_{\text{eff}}}{p^\ast}
\left(\dfrac{k_{\text{eff}}-p k }{k_{\text{eff}}(k - k_{\text{eff}})} \right)
\label{eq_emt_1}
\end{align}
where $p^{\ast}=2d/z$ and which is now in a suitable form to be used for the numerical evaluation of $G_{11}$ \cite{Garboczi1985}.
For the present purposes, we note that the Green's function $G_{11}$ is given by 
\begin{align}
G_{11}(\omega^2,k_{\text{eff}})
=
\dfrac{1}{2m N}
\sum_{\underline{k},i} \dfrac{1}{\omega^2-\omega^2_i(\underline{q})}
\label{eq_emt_2}
\end{align}
where the $\underline{q}$-sum runs over the first Brillouin zone and $i$-sum over the branches of the dispersion relation of the underlying crystal obtained from as the eigenfunctions of the corresponding dynamical matrix.
Finally the vibrational density $D_{\text{CPA}}$ of states is obtained by the relation~\cite{Garboczi1985}
\begin{align}
D_{\text{CPA}}(\omega^2,k_{\text{eff}}) =- \dfrac{1}{\pi} \text{Im} G_{11}(\omega^2,k_{\text{eff}}).
\end{align}

\subsection{CPA for the 3D FCC}
In order to quantitatively study the effective medium theory in the case for the 3D FCC we need to write down the dynamical matrix for the perfect FCC and determine the eigenfunctions $\omega_i^2(\underline{k})$.
To keep the notation clean we set the spring constant, mass and lattice spacing to 1, i.e. $k=m=a=1$. The resulting $3 \times 3$ dynamical matrix $\mathcal{D}$ is given in terms of its components
by
\begin{align*}
\mathcal{D}_{11}=&-2 \left(\cos \left(\frac{q_x}{2}\right) \left(\cos
   \left(\frac{q_y}{2}\right)+\cos
   \left(\frac{q_z}{2}\right)\right)-2\right)
\\
\mathcal{D}_{22}=&-2 \left(\cos \left(\frac{q_y}{2}\right) \left(\cos
   \left(\frac{q_x}{2}\right)+\cos
   \left(\frac{q_z}{2}\right)\right)-2\right)
\\
\mathcal{D}_{33}=&4-2 \cos \left(\frac{q_z}{2}\right) \left(\cos
   \left(\frac{q_x}{2}\right)+\cos \left(\frac{q_y}{2}\right)\right)
\\
\mathcal{D}_{12}=&\mathcal{D}_{21}=2 \sin \left(\frac{q_x}{2}\right) \sin \left(\frac{q_y}{2}\right)
\\
\mathcal{D}_{13}=&\mathcal{D}_{31}=2 \sin \left(\frac{q_x}{2}\right) \sin \left(\frac{q_z}{2}\right)
\\
\mathcal{D}_{23}=&\mathcal{D}_{32}=2 \sin \left(\frac{q_y}{2}\right) \sin \left(\frac{q_z}{2}\right).
\end{align*}
By diagonalizing $\mathcal{D}$ we can compute the three branches  $\omega_i^2(\underline{q}, k)$ of the dispersion relation. In order to solve the EMT equations \eqref{eq_emt_1} and \eqref{eq_emt_2} we replace the spring constant $k$ in the dispersion relation with the spring constant of the effective medium $k_{\text{eff}}$. Since we need to solve \eqref{eq_emt_1} and \eqref{eq_emt_2} iteratively, we use  $k_{\text{eff}}=1+i$ as a starting value. Then we compute  the effective-medium Green's function$G_{11}(\omega^2,k_{\text{eff}}=1+i)$ by numerically summing the three branches of the dispersion relation over $10^5$ points in the first Brillioun zone of the fcc crystal. The resulting Green's function is then substituted in Eq.~ \eqref{eq_emt_1} to obtain the new value for the effective spring constant $k_{\text{eff}}$. This process is repeated until convergence of $k_{\text{eff}}$ reached, which happens after about 10 iterations.

\subsection{Inversion-symmetry breaking parameter}
	Here we present an argument, why the normalisation of the ISB parameter is not as arbitrary as it seems, but quite reasonable.
	\\
	We start from a system with an arbitrary distribution of angles $\theta$ and $\phi$ (here they are the angles of the bonds with respect to some reference system - not the angles between bonds in the system). The only thing we can say in the framework of the affine force field is that for every bond vector $\hat{n}_{i  j}$ exists a vector $\hat{n}_{j  i} = - \hat{n}_{i  j}$ with the same relative frequency (here frequency refers to the probability and not to some oscillation). We now write the general expression of $|\underline{\Xi}|^2$.
	\begin{align}
	|\underline{\Xi}|^2 \,=\, \kappa^2 r_0^2 \sum_{i} \sum_{\alpha }\left( \sum_{j\, nn\, i} \hat{n}_{i j}^\alpha \hat{n}_{i j}^x \hat{n}_{i j}^y \right)^2
	\end{align}
	where $\alpha = x,y,z$ are the Cartesian coordinates. We can carry out those sums and regroup the terms to get
	\begin{align}\label{AFexp}
	|\underline{\Xi}|^2 \,&=\, \kappa^2 r_0^2 \left(\sum_{i j} \left(\hat{n}_{i j}^x \hat{n}_{i j}^y \right)^2\right.  \notag\\&+ \left.\sum_{i} \sum_{k, l \, nn\, i} (\hat{n}_{i k} \cdot \hat{n}_{i l})(\hat{n}_{i k} \cdot \hat{n}_{i l})^x (\hat{n}_{i k} \cdot \hat{n}_{i l})^y \right)
	\end{align}
	Now we implement the difference between the most random configuration, which we call isotropic in the case of the random network, and any other configuration that we want to calculate the ISB parameter for. \\
	With no further restrictions, the second term in \eqref{AFexp} is zero. We can explain this by the fact that, as said before, the probability to have any vector according to a given angel distribution is equal to the probability to have the negative of this vector. In the framework of the scalar product this means that 
	\begin{align}\label{AFav}
		&P((\hat{n}_{i k} \cdot \hat{n}_{i l})(\hat{n}_{i k} \cdot \hat{n}_{i l})^x (\hat{n}_{i k} \cdot \hat{n}_{i l})^y)\notag \,\\ \notag & \qquad \qquad \quad =\\ \notag &P(-(\hat{n}_{i k} \cdot \hat{n}_{i l})(\hat{n}_{i k} \cdot \hat{n}_{i l})^x (\hat{n}_{i k} \cdot \hat{n}_{i l})^y)\\
		&\longrightarrow \left\langle 	(\hat{n}_{i k} \cdot \hat{n}_{i l})(\hat{n}_{i k} \cdot \hat{n}_{i l})^x (\hat{n}_{i k} \cdot \hat{n}_{i l})^y \right\rangle \,=\, 0
	\end{align}
	In a hard sphere system you would have the restriction that $\hat{n}_{i k} \cdot \hat{n}_{i l} < 0.5$ since two bonds from an atom $i$ cannot have an angle smaller that $\pi/3$. This shifts the average in \eqref{AFav} to a negative value and lowers $|\underline{\Xi}|^2$. This also is the core mechanic of our ISB parameter. So what remains of $|\underline{\Xi}|^2$ in the total random case is
	\begin{equation}\label{AFiso}
	|\underline{\Xi}|^2_{random} \,=\, \kappa^2 r_0^2 \sum_{i j} \left(\hat{n}_{i j}^x \hat{n}_{i j}^y \right)^2.
	\end{equation}
	So the ISB parameter becomes
	\begin{equation}\label{OrderParameter}
	F_{IS}\,=\,1\,-\,\frac{|\underline{\Xi}|^2}{\kappa^2 r_0^2 \sum_{i j} \left(\hat{n}_{i j}^x \hat{n}_{i j}^y \right)^2}
	\end{equation}
	which reproduces the right behaviour of the ISB parameter. When we combine \eqref{OrderParameter} and \eqref{AFexp} we can get an even further simplified expression
	\begin{equation}\label{OrderParameterSim}
	\begin{gathered}
	F_{IS}\,=\,-\frac{\sum_{i} \sum_{k, l \, nn\, i} (\hat{n}_{i k} \cdot \hat{n}_{i l})(\hat{n}_{i k} \cdot \hat{n}_{i l})^x (\hat{n}_{i k} \cdot \hat{n}_{i l})^y}{\sum_{i j} \left(\hat{n}_{i j}^x \hat{n}_{i j}^y \right)^2}\\
	=\, -\frac{\sum_{i} \sum_{k, l \, nn\, i} \,\cos \alpha_{k l}\,(\hat{n}_{i k} \cdot \hat{n}_{i l})^x (\hat{n}_{i k} \cdot \hat{n}_{i l})^y}{\sum_{i j} \left(\hat{n}_{i j}^x \hat{n}_{i j}^y \right)^2}
	\end{gathered}
	\end{equation}
	Where we have linked our inversion symmetry breaking parameter to the angular distribution of the angles $\alpha_{k l}$ between bonds in each cell of the system.\\
	But one problem remains: We have defined our ISB parameter in the framework of $xy$ shearing. So we weighted the symmetry in the $xy$ plane higher than the symmetry in the other directions. To get a general parameter we have to include the other directions, represented by their corresponding affine force fields, as well. So we replace $|\underline{\Xi}|^2 = \sum_{a,b = x,y,z} |\underline{\Xi}_{a b}|^2$. Therefore equation \eqref{AFiso} and the numerator of \eqref{OrderParameterSim} becomes
	\begin{align*}\label{AFisoGen}
	&|\underline{\Xi}|^2_{\text{random}} \,=\, \kappa^2 r_0^2 \sum_{i j} \sum_{a b} \left(\hat{n}_{i j}^a \hat{n}_{i j}^b \right)^2\, \\ \notag &=\, \kappa^2 r_0^2 \sum_{i j} \left(\left(\hat{n}_{i j}^x\right)^2 + \left(\hat{n}_{i j}^y\right)^2 + \left(\hat{n}_{i j}^z\right)^2 \right)^2 \\
	\,&=\, N Z \kappa^2 r_0^2
	\sum_{i} \sum_{k, l \, nn\, i} \sum_{a b} (\hat{n}_{i k} \cdot \hat{n}_{i l})(\hat{n}_{i k} \cdot \hat{n}_{i l})^a (\hat{n}_{i k} \cdot \hat{n}_{i l})^b \\
	 \,&=\, \sum_{i} \sum_{k, l \, nn\, i} (\hat{n}_{i k} \cdot \hat{n}_{i l})\left( (\hat{n}_{i k} \cdot \hat{n}_{i l})^x + (\hat{n}_{i k} \cdot \hat{n}_{i l})^y + (\hat{n}_{i k} \cdot \hat{n}_{i l})^z \right)^2\\
	&\,=\, \sum_{i} \sum_{k, l \, nn\, i} (\hat{n}_{i k} \cdot \hat{n}_{i l})^3
	\end{align*}	
	The ISB parameter therefore becomes 
	\begin{equation}\label{OrderParameterGen}
	\begin{gathered}
	F_{IS}\,=\,-\frac{1}{N Z} \sum_{i} \sum_{k, l \, nn\, i} (\hat{n}_{i k} \cdot \hat{n}_{i l})^3 \,=\,-\frac{1}{N Z} \sum_{i} \sum_{k, l \, nn\, i} (\cos \alpha_{k l})^3
	\end{gathered}
	\end{equation}
	It is important to notice that we count each angle twice. Due to the restriction $\cos \alpha_{k l} < 0.5$, the value $\left\langle (\cos \alpha_{k l})^3 \right\rangle$ is smaller than $1$. So the sign of \eqref{OrderParameterGen} is correct to produce a parameter $F_{\text{IS}} < 1$.

\balance

\newpage
\bibliographystyle{rsc} 
\providecommand*{\mcitethebibliography}{\thebibliography}
\csname @ifundefined\endcsname{endmcitethebibliography}
{\let\endmcitethebibliography\endthebibliography}{}

\end{document}